\documentclass[prd,twocolumn,showpacs,superscriptaddress,nofootinbib,floatfix,showkeys,10pt]{revtex4-1}

\usepackage{graphicx}
\usepackage{amsmath}
\usepackage{bm}
\usepackage{yhmath}
\usepackage{mathtools}
\usepackage{wasysym}
\usepackage[colorlinks,citecolor=blue,urlcolor=blue,linkcolor=blue]{hyperref}
\usepackage{subfigure}
\usepackage{color}
\usepackage{cases}
\usepackage{subfigure}
\usepackage{times}
\usepackage{dcolumn,booktabs,bm}
\usepackage{slashed}
\usepackage{amsfonts,amssymb,stmaryrd,latexsym}
\usepackage{mathrsfs}
\usepackage{textcomp}
\usepackage{multirow}
\usepackage{cancel}
\usepackage{array}
\usepackage{enumerate,enumitem}
\usepackage{orcidlink}
\usepackage{dashrule}
\usepackage{multirow}
\usepackage{times}

\allowdisplaybreaks[4]
\newcounter{cnt}
\makeatletter
\let\oldhypertarget\hypertarget
\renewcommand{\hypertarget}[2]{%
  \oldhypertarget{#1}{#2}%
    \protected@write\@mainaux{}{%
        \string\expandafter\string\gdef
          \string\csname\string\detokenize{#1}\string\endcsname{#2}%
    }%
  }
\newcommand{\myhyperlink}[1]{%
  \hyperlink{#1}{\csname #1\endcsname}%
  }
\makeatother

\def\N2LO{{N$^2$LO}}
\newcommand{\etal}{\textit{et al}.}

\begin{document}


\title{Spectrum of the molecular tetraquarks: Unraveling the $T_{cs0}(2900)$ and $T_{c\bar{s}0}^a(2900)$}
\author{Bo Wang\,\orcidlink{0000-0003-0985-2958}}
\affiliation{College of Physics Science \& Technology, Hebei University, Baoding 071002, China}
\affiliation{Hebei Key Laboratory of High-precision Computation and Application of Quantum Field Theory, Baoding, 071002, China}
\affiliation{Hebei Research Center of the Basic Discipline for Computational Physics, Baoding, 071002, China}

\author{Kan Chen\,\orcidlink{0000-0002-1435-6564}}\email{chenk10@nwu.edu.cn}
\affiliation{School of Physics, Northwest University, Xi’an 710127, China}

\author{Lu Meng\,\orcidlink{0000-0001-9791-7138}}\email{lu.meng@rub.de}
\affiliation{ Institut f\"ur Theoretische Physik II, Ruhr-Universit\"at Bochum,  D-44780 Bochum, Germany}

\author{Shi-Lin Zhu\,\orcidlink{0000-0002-4055-6906}}\email{zhusl@pku.edu.cn}
\affiliation{School of Physics and Center of High Energy Physics, Peking University, Beijing 100871, China}

\begin{abstract}
We relate the interactions of the $\bar{D}^{(\ast)} K^\ast$ and $D^{(\ast)} K^\ast$ systems to those of $D^{(\ast)}D^{(\ast)}$ and $D^{(\ast)}\bar{D}^{(\ast)}$ respectively,  considering the residual strong interactions at the near-threshold energy is too weak to excite the strange quarks inside the hadrons. We propose an effective model to describe the low-energy S-wave interactions that are undertaken by the light $u$, $d$ quarks between two separated heavy hadrons. We find that the existence of molecules in the heavy-(anti)heavy sectors will naturally lead to the emergence of molecular states in $\bar{D}^{(\ast)} K^\ast$ and $D^{(\ast)} K^\ast$ systems. The recently observed $T_{cs0}(2900)$ and $T_{c\bar{s}0}^a(2900)$ can be well identified as the $0(0^+)$ and $1(0^+)$ partners of $T_{cc}(3875)$ and $Z_c(3900)$ in the charmed strange sector, respectively. We also predict their members under the {\it heavy} ($c$ and $s$) quark symmetry and SU(2) flavor symmetry. Most of them are very good molecule candidates, for example, (i) the $0(1^+)$ states in $D^\ast D^\ast$, $\bar{D}K^\ast$, $\bar{D}^\ast K^\ast$; (ii) the $0^{(+)}(2^{+(+)})$ states in $D^\ast \bar{D}^\ast$, $\bar{D}^\ast K^\ast$, $D^\ast K^\ast$; (iii) the $1^-(0^{++})$ state in $D^\ast\bar{D}^\ast$ and $1(1^+)$ state in $D^\ast K^\ast$. The $0^+(0^{++})$ state in $D\bar{D}$ and the $0(1^+)$ state in $DK^\ast$ might also exist as virtual states, and the $0(1^+)$ $DK^\ast$ can serve as a key to infer the existence of $0^+(0^{++})$ $D\bar{D}$. The $D_s\pi$ invariant mass spectrum of $T_{c\bar{s}0}^a(2900)$ is also studied within the coupled-channel approach, and the molecular interpretation of $T_{c\bar{s}0}^a(2900)$ is consistent with the experimental data. Searching for the predicted states in experiments is crucial to discriminate the different pictures for interpreting these near-threshold exotica.
\end{abstract}

\maketitle

\section{Introduction}\label{sec:intro}

Over the last two decades, experiments have unveiled numerous near-threshold exotic hadron states, indicating novel quantum chromodynamics (QCD) structures in the heavy flavor sector~\cite{Workman:2022ynf}. These states are considered exotic due to their inability to align with known meson and baryon configurations in the conventional quark model. Various interpretations have been proposed to comprehend these experimental observations. These interpretations encompass hadronic molecules, compact multiquark states, and kinematic effects, etc. Among these, the hadronic molecular explanation takes precedence due to their commonly observed near-threshold features (e.g., see the reviews~\cite{Chen:2016qju,Guo:2017jvc,Liu:2019zoy,Lebed:2016hpi,Esposito:2016noz,Brambilla:2019esw,Chen:2021ftn,Chen:2022asf,Meng:2022ozq}).

The first as well as the most famous exotic state is the isoscalar $X(3872)$ in the charmonium energy region, which was observed by the Belle Collaboration in 2003~\cite{Belle:2003nnu}. Its mass is extremely close to the $D^0\bar{D}^{\ast0}$ threshold, e.g., $m_{D^0\bar{D}^{\ast0}}-m_{X(3872)}=0.04\pm0.12$ MeV~\cite{Workman:2022ynf}. The isovector states $Z_c(3900)$ and $Z_c(4020)$ were observed by the BESIII Collaboration~\cite{BESIII:2013ris,BESIII:2013ouc}, and their masses are close to the $D\bar{D}^\ast$ and $D^\ast\bar{D}^\ast$ thresholds, respectively~\cite{Workman:2022ynf}. The LHCb Collaboration observed three hidden-charm pentaquark states, the $P_c(4312)$, $P_c(4440)$, and $P_c(4457)$~\cite{LHCb:2019kea}. They lie below the $\Sigma_c\bar{D}$ and $\Sigma_c\bar{D}^\ast$ thresholds about several to tens MeVs, respectively~\cite{Workman:2022ynf}. The LHCb Collaboration also observed a double-charm tetraquark state $T_{cc}(3875)$. Its mass is below the $D^0D^{\ast+}$ threshold about $300$ keV~\cite{LHCb:2021vvq,LHCb:2021auc}.

Recently, the LHCb Collaboration observed two charmed strange tetraquarks $T_{cs0}(2900)$ and $T_{cs1}(2900)$ in the $D^-K^+$ invariant mass distributions from the decay $B^+\to D^+D^-K^+$~\cite{LHCb:2020bls,LHCb:2020pxc} (we will refer to the new naming scheme for the exotic hadrons according to their quark contents, isospins, and spin-parities~\cite{Gershon:2022xnn}).  Their $J^P$ quantum numbers were undetermined yet, but the $0^+$ and $1^-$ are preferred for the $T_{cs0}(2900)$ and $T_{cs1}(2900)$, respectively. The $1^-$ assignment for $T_{cs1}(2900)$ is necessary because evident P-wave behavior is observed in the $D^-K^+$ helicity angle distributions~\cite{LHCb:2020pxc}. The closeness of their masses to the $\bar{D}^\ast K^\ast$ threshold has prompted molecular interpretations~\cite{Chen:2020aos,He:2020btl,Liu:2020nil,Hu:2020mxp,Agaev:2020nrc,Wang:2021lwy} (the $\bar{D}^\ast K^\ast$ molecules had been predicted before the experimental observations~\cite{Molina:2010tx}). In addition, other pictures such as the compact $\bar{c}\bar{s}ud$ tetraquarks~\cite{Karliner:2020vsi,He:2020jna,Wang:2020xyc,Zhang:2020oze,Wang:2020prk,Lu:2020qmp,Tan:2020cpu,Albuquerque:2020ugi} and the kinematic effects from the triangle singularities~\cite{Liu:2020orv,Burns:2020epm} are also devoted to understanding the mass spectra and the lineshapes of $T_{cs0,1}(2900)$, respectively. The calculations from the quark model~\cite{Wang:2020prk,Lu:2020qmp} showed that the mass of an isoscalar S-wave compact tetraquark $\bar{c}\bar{s}ud$ is lower than the experimental mass of the $T_{cs0}(2900)$. Therefore, the compact tetraquark explanation appears incompatible with the current experimental data. In Ref.~\cite{Wang:2021lwy}, we studied the $\bar{D}^\ast K^\ast$ interactions within the spirit of chiral effective field theory and fitted the $D^-K^+$ invariant mass distributions, in which the $T_{cs0}(2900)$ and $T_{cs1}(2900)$ can be interpreted as the isoscalar S-wave and P-wave excitation of the $\bar{D}^\ast K^\ast$ bound states, respectively. Various aspects of the $T_{cs}$ states were investigated in literature, such as the strong decays~\cite{Huang:2020ptc,Xiao:2020ltm}, the branching fractions from $B$ decays~\cite{Chen:2020eyu}, the lineshapes~\cite{Chen:2021tad}, the productions through the kaon-induced reactions~\cite{Lin:2022eau}, and the searching for its $1^+$ and $2^+$ partners~\cite{Dai:2022htx,Bayar:2022wbx}, and so on.

Very recently, the LHCb also observed a doubly charged tetraquark $T_{c\bar{s}0}^a(2900)^{++}$ and its neutral partner $T_{c\bar{s}0}^a(2900)^{0}$ within the $D_s^+\pi^+$ and $D_s^+\pi^-$ invariant mass distributions in the $B$ decay processes $B^+\to D^-D_s^+\pi^+$ and $B^0\to\bar{D}^0D_s^+\pi^-$, respectively~\cite{LHCb:2022sfr,LHCb:2022lzp}. The spin-parity $J^P$ is determined to be $0^+$. These states were predicted in Ref. \cite{Chen:2017rhl}. 
Given their comparable masses and widths, these two states should belong to the same isospin triplet. Like the $T_{cs0}(2900)$, the $T_{c\bar{s}0}^a(2900)$  is also situated near the $D^\ast K^\ast$ threshold. Thus, some works interpret it as the $D^\ast K^\ast$ hadronic molecule~\cite{Chen:2022svh,Agaev:2022eyk,Yue:2022mnf,Duan:2023lcj}, while Ref.~\cite{Ke:2022ocs} argues against a binding solution in the isovector $D^\ast K^\ast$ system. The compact $c\bar{s}u\bar{d},c\bar{s}\bar{u}d$ tetraquark explanation was proposed in Refs.~\cite{Liu:2022hbk,Yang:2023evp,Lian:2023cgs,Wei:2022wtr,Ortega:2023azl}. The authors in Ref.~\cite{Molina:2022jcd} reproduced the experimental peak in $D_s^+\pi^+$ channel as a threshold cusp arising from the $D^\ast K^\ast$$-$$D_s^\ast \rho$ coupled-channel interactions. In Ref.~\cite{Yue:2022mnf}, it was shown that in the molecular picture, the dominant decay mode of $T_{c\bar{s}0}^a(2900)$ is $DK$ rather than the experimentally observed channel $D_s\pi$. Thus the authors of Ref.~\cite{Duan:2023qsg} suggest to search for the $T_{c\bar{s}0}^a(2900)^{++}$ also in the $B^+\to K^+D^+D^-$ process. The discovery of $T_{c\bar{s}0}^a(2900)$ inspired the studies on its kaon-induced productions~\cite{Huang:2023fvj}, and the charmed strange pentaquarks as well~\cite{An:2022vtg,Chen:2023qlx}.

From the $X(3872)$ to $T_{c\bar{s}0}^a(2900)$, we have reviewed some of the well-established states that have emerged from experiments over the past two decades. We mainly focused on the recently observed $T_{cs}$ and $T_{c\bar{s}}$ states. Can we develop a unified framework to describe these states, considering their distinctive characteristics that are close to their corresponding thresholds? In this study, we will combine symmetry analyses with dynamic calculations to investigate the aforementioned exotic states. We aim to demonstrate that the presence of $T_{cs0}(2900)$ and $T_{c\bar{s}0}^a(2900)$ as $\bar{D}^\ast K^\ast$ and $D^\ast K^\ast$ molecules, respectively, naturally emerges due to the SU(2) flavor symmetry and the {\it heavy} quark symmetry at the hadron level.

A hadronic molecule is commonly defined as an entity composed of two color-neutral hadrons. A well-known example is the deuteron, a bound state of two nucleons with a binding energy of approximately $2.2$ MeV. This concept was applied broadly rather than being exclusive to nucleon-nucleon systems. For instance, two heavy hadrons with specific quantum numbers can also cluster together, resulting in the formation of molecular states due to van der Waals-like residual strong interactions. This interaction between color-neutral objects is significantly weaker compared to the quark-gluon couplings within the hadrons. This characteristic can naturally elucidate why the mentioned molecular candidates closely approach the thresholds of paired heavy hadrons. In practice, the observed molecular candidates are typically located near the nearest thresholds, often within a few MeVs' range.

The binding energies $E_b$ of the heavy hadronic molecules is at the order of $1$ MeV, e.g., the $X(3872)$~\cite{Workman:2022ynf} and $T_{cc}(3875)$~\cite{LHCb:2021auc}. Thus, the typical scale of the binding momentum is $\gamma_b^{\text{ty}}=\sqrt{2\mu E_b}\lesssim 100$ MeV (with $\mu$ the reduced mass of $DD^\ast$).  The typical strange quark mass in the constitute quark model is about $m^{\text{QM}}_s\sim 500$ MeV~\cite{Silvestre-Brac:1996myf,Vijande:2004he}. One sees that the $\gamma_b^{\text{ty}}\ll m^{\text{QM}}_s$, i.e., the $m^{\text{QM}}_s$ can be regarded as a hard scale\footnote{An extra  energy scale could arise if we consider the mass splitting of $D$ and $D^\ast$, denoted as $\delta^\prime=m_{D^\ast}-m_D\approx140$ MeV~\cite{Meng:2020cbk,Meng:2021rdg}. This of course is another hard scale with typical momentum $p^{\prime\text{ty}}=\sqrt{2\mu \delta^\prime}\sim 520$ MeV. So the coupled-channel contributions, e.g., the $DD^\ast$$-$$D^\ast D^\ast$ coupling will not be incorporated into our calculations. We will primarily focus on the single-channel dynamics.}, which implies that the near-threshold interactions are too weak to excite the strange quarks inside the heavy hadrons. In other words, the strange quark will behave like an inert source ({\it heavy} quark) from the view of the residual strong interactions at the near-threshold energy scale. Therefore, we will call the strange and charm quarks together as the {\it heavy} quarks for the near-threshold interactions in what follows\footnote{The above argument is valid for the {\it heavy} vector $K^\ast$ mesons ($m_{K^\ast}\sim m_N$) rather than the pseudoscalar kaons due to the non-negligible Goldstone nature of the kaon. So we will not consider the $D_{s0}^\ast(2317)$ and $D_{s1}(2460)$ in this work because of the entanglement with the $DK$ and $D^\ast K$ correlations, respectively.}, and we will use the SU(2) flavor symmetry instead of the SU(3) in the calculations.

\begin{figure}[!ht]
\begin{centering}
    \scalebox{1.0}{\includegraphics[width=\linewidth]{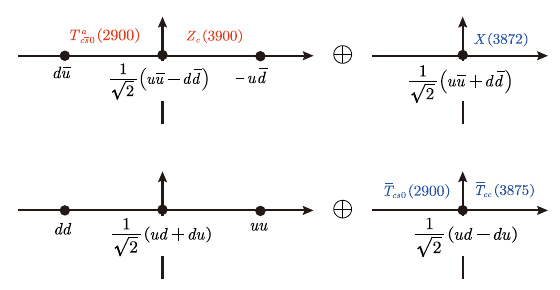}}
    \caption{The isospin triplet (left column) and singlet (right column) of the di-meson systems that contain a light $q\bar{q}$ (first row) or $qq$ (second row) pair in the SU(2) flavor symmetry, respectively.  \iffalse The corresponding molecular candidates from the $D^\ast K^\ast$ ($\bar{D}^\ast K^\ast$) and $D\bar{D}^\ast$ ($DD^\ast$) systems are placed at the left and right side of the $y$-axis, respectively, while the positions in blank shall correspond to our predictions in Sec.~\ref{sec:spec}. \fi \label{fig:flavor}}
\end{centering}
\end{figure}

Based on the aforementioned analyses, a direct connection can be established between $D^{(\ast)}K^\ast~(\bar{D}^{(\ast)}K^\ast)$ and $D^{(\ast)}\bar{D}^\ast~(\bar{D}^{(\ast)}\bar{D}^\ast)$ systems. In other words, the recently observed $T_{c\bar{s}0}^a(2900)$ and $T_{cs0}(2900)$ could potentially represent the charmed strange counterparts of $Z_c(3900)$ and $T_{cc}(3875)$, respectively, e.g., see the Fig.~\ref{fig:flavor}. Similar to the charm quark, the strange quark also has limited involvement in the near-threshold interactions. Consequently, the interactions within the paired $D^{(\ast)}K^\ast$, $\bar{D}^{(\ast)}K^\ast$, $D^{(\ast)}\bar{D}^\ast$, and $\bar{D}^{(\ast)}\bar{D}^\ast$ systems are primarily mediated by the interactions between the light $u$ and $d$ quarks.  Once we get $V_{\bar{q}q}$ and $V_{qq}$, we can proceed to analyze the associated hadronic molecule candidates as shown in Table~\ref{tab:qc}. 

This paper is organized as follows. An effective model for the S-wave near-threshold interactions of two {\it heavy} hadrons is illustrated in Sec.~\ref{sec:model}. The determination of the low-energy constants, the molecule spectra of considered systems and related discussions are presented in Sec.~\ref{sec:nums}. The $D_s\pi$ invariant mass distributions of $T_{c\bar{s}0}^a(2900)$ is studied in Sec.~\ref{sec:linesh}. The summary and outlook are given in Sec.~\ref{sec:sum}.

\begin{table}[!ht]
\centering
\renewcommand{\arraystretch}{1.3}
\caption{The molecule candidates and their corresponding nearest thresholds, the quark contents, and the interaction types that govern the formation of the hadronic molecules. The $P_c$s denote the three hidden-charm pentaquarks $P_c(4312)$, $P_c(4440)$ and $P_c(4457)$.\label{tab:qc}}
\setlength{\tabcolsep}{2.8mm}
{
\begin{tabular}{cccc}
\toprule[0.8pt]
States & Nearest thresholds & Quark contents & $V_{\textrm{type}}$\tabularnewline
\hline
$X(3872)$ & $D\bar{D}^{\ast}$ & $[c\bar{q}][\bar{c}q]$ & \multirow{3}{*}{$V_{\bar{q}q}$}\\
$Z_{c}(3900)$ & $D\bar{D}^{\ast}$ & $[c\bar{q}][\bar{c}q]$ \\
$T_{c\bar{s}0}^{a}(2900)$ & $D^{\ast}K^{\ast}$ & $[c\bar{q}][\bar{s}q]$ \\
\hline
$P_{c}$s & $\Sigma_{c}\bar{D}^{(\ast)}$ & $[cqq][\bar{c}q]$ & \multirow{3}{*}{$V_{qq}$}\\
$\bar{T}_{cc}(3875)$ & $\bar{D}\bar{D}^{\ast}$ & $[\bar{c}q][\bar{c}q]$\\
$\bar{T}_{cs0}(2900)$ & $\bar{D}^{\ast}K^{\ast}$ & $[\bar{c}q][\bar{s}q]$\\
\bottomrule[0.8pt]
\end{tabular}

}
\end{table}

\section{Near-threshold effective potentials for the S-wave interactions}\label{sec:model}

As we have elucidated in Sec.~\ref{sec:intro}, one can obtain the following correspondence for the near-threshold interactions,
\begin{eqnarray}\label{eq:corresp}
\bar{D}^{(\ast)}K^\ast \longleftrightarrow \bar{D}^{(\ast)}\bar{D}^\ast, \qquad  D^{(\ast)}K^\ast \longleftrightarrow D^{(\ast)}\bar{D}^\ast,
\end{eqnarray}
since the two systems arise from the same $qq$ or $\bar{q}q$ interactions, respectively. In the following, we will illustrate the low-energy forms of $V_{qq}$ and $V_{\bar{q}q}$.

\begin{figure}[!ht]
\begin{centering}
    \scalebox{0.8}{\includegraphics[width=\linewidth]{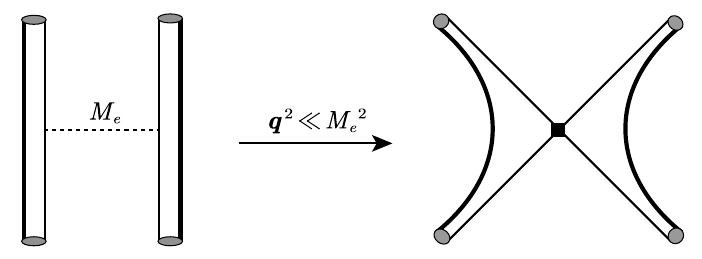}}
    \caption{The schematic diagrams for the near-threshold interactions that are undertaken by the light quark interplays. The thick, thin and dashed lines denote the {\it heavy} quarks ($c$ and $s$), light quarks ($u$ and $d$) and the exchanged mesons (isospin triplet plus the singlet), respectively.\label{fig:feyndiagram}}
\end{centering}
\end{figure}

The residual strong interactions between two {\it heavy}-light hadrons can be described using the one-boson exchange model. The theoretically allowed bosons could be the pseudoscalar ($\pi$, $\eta$, $\dots$), scalar ($a_0$, $f_0$, $\dots$), vector ($\rho$, $\omega$, $\dots$), axial-vector ($a_1$, $f_1$, $\dots$), and tensor ($a_2$, $f_2$, $\dots$), etc. Currently, determining the specific couplings of {\it heavy}-light hadrons with these potentially exchanged bosons is a challenging task. Consequently, our approach in this work is to consider their cumulative contributions. Here, we will adopt the concept of the non-relativistic chiral quark model~\cite{Manohar:1983md}, assuming that the exchange interactions occur at the quark level. This allows us to construct effective Lagrangians and compute non-relativistic effective potentials between the light quarks, e.g., see the vector couplings in Ref.~\cite{Riska:2000gd}. Anyhow, the non-relativistic effective potentials should have the following form
\begin{eqnarray}\label{eq:Veff}
V_{\text{eff}}&\sim&\sum_{e}\frac{\{1,\bm{\sigma}_1\cdot\bm{\sigma}_2,(\bm{\sigma}_1\cdot\bm{q})(\bm{\sigma}_2\cdot\bm{q}),\dots\}}{\bm{q}^2+M_e^2},
\end{eqnarray}
where the $\bm q$ and $M_e$ represent the transferred momenta and the masses of the exchanged mesons, respectively. The corresponding form of the operators in the numerator depends on the quantum numbers of the exchanged particles. For the near-threshold interactions ($\bm{q}^2\ll M_e^2$), one can reduce the non-local interaction to the local form by keepig the leading term in the following expansion
\begin{eqnarray}\label{eq:Vexpd}
\frac{1}{\bm{q}^2+M_e^2}=\frac{1}{M_e^2}\left(1-\frac{\bm{q}^2}{M_e^2}+\dots\right),
\end{eqnarray}
see also the schematic diagrams in Fig.~\ref{fig:feyndiagram}.
The next-to-leading order contributions can be safely neglected due to the suppression of $\bm{q}^2/M_e^2$. Thus, only the momentum-independent central term and spin-spin interaction in Eq.~\eqref{eq:Veff} appear in the leading terms in calculations.


In the light quark ($u$ and $d$) sector, the exchanged meson fields contain both the isospin triplet and singlet.  We take the scalar field $\mathscr{S}$ as an example and assume it can be written as
\begin{eqnarray}
\mathscr{S}	&=&	\mathscr{S}_{i}\tau_{i}+\frac{1}{\sqrt{2}}\mathscr{S}_{1},
\end{eqnarray}
where $\tau_i$ is the Pauli matrices in the isospin space, and $\mathscr{S}_i$ and $\mathscr{S}_1$ denote the isospin triplet and singlet, respectively. Therefore, the net contributions from the exchanged particles will give rise to the non-relativistic effective potentials for the light $qq$ and $\bar{q}q$ with the following forms, respectively,
\begin{eqnarray}
V_{qq}&=&\left(\bm{\tau}_{1}\cdot\bm{\tau}_{2}+\frac{1}{2}\tau_{0,1}\cdot\tau_{0,2}\right)(c_{s}+c_{a}\bm{\sigma}_{1}\cdot\bm{\sigma}_{2}),\label{eq:Vqq}\\
V_{\bar{q}q}&=&\left(-\bm{\tau}_{1}^\ast\cdot\bm{\tau}_{2}+\frac{1}{2}\tau_{0,1}\cdot\tau_{0,2}\right)(\tilde{c}_{s}+\tilde{c}_{a}\bm{\sigma}_{1}\cdot\bm{\sigma}_{2}),\label{eq:Vqqbar}
\end{eqnarray}
in which the $\tau_0$ denotes the $2\times2$ identity matrix.  The terms $c_s$ ($\tilde{c}_s$) and $c_a$ ($\tilde{c}_a$) denote the low-energy constants (LECs), signifying the magnitude of the central potential and spin-spin couplings, respectively.

For the later convenience, the operators in Eqs.~\eqref{eq:Vqq} and \eqref{eq:Vqqbar} will be decomposed into four parts (e.g., for the $qq$ case),
\begin{align}
\mathcal{O}_1&=\frac{1}{2}\tau_{0,1}\cdot\tau_{0,2},& \mathcal{O}_2&=\bm{\tau}_{1}\cdot\bm{\tau}_{2},\nonumber\\
\mathcal{O}_3&=\frac{1}{2}\tau_{0,1}\cdot\tau_{0,2}\bm{\sigma}_{1}\cdot\bm{\sigma}_{2},& \mathcal{O}_4&=\bm{\tau}_{1}\cdot\bm{\tau}_{2}\bm{\sigma}_{1}\cdot\bm{\sigma}_{2},
\end{align}
in which the $\bm{\tau}_1$ will be replaced with $-\bm{\tau}_1^\ast$ for the $\bar{q}q$ case. The matrix elements of the operators $\mathcal{O}_i$ are given as
\begin{eqnarray}\label{eq:matOi}
\langle\mathcal{O}_i\rangle=\left\langle [\mathscr{H}_1\mathscr{H}_2]_J^I\left|\mathcal{O}_i\right|[\mathscr{H}_1\mathscr{H}_2]_J^I \right\rangle,
\end{eqnarray}
where the $[\mathscr{H}_1\mathscr{H}_2]_J^I$ denotes the flavor and spin wave function of the di-hadron system with the specified total angular momentum $J$ and total isospin $I$, and it can be expressed as the direct product of the flavor and spin wave functions of the individual $\mathscr{H}_1$ and $\mathscr{H}_2$. One can see more details in Refs.~\cite{Chen:2021cfl,Chen:2021spf}. The hadron-level potentials of di-hadron systems can be obtained via (e.g., for the $qq$ case)
\begin{eqnarray}\label{eq:Vsys}
V_{\text{sys}}&=&c_{s}\left[\langle\mathcal{O}_{1}\rangle+\langle\mathcal{O}_{2}\rangle\right]+c_{a}\left[\langle\mathcal{O}_{3}\rangle+\langle\mathcal{O}_{4}\rangle\right].
\end{eqnarray}
Translating the $V_{\bar{q}q}$ into the hadron-level potentials $\tilde{V}_{\text{sys}}$ takes the similar form as those in Eqs.~\eqref{eq:matOi} and \eqref{eq:Vsys}.

Here, we need to emphasize that deriving the hadron-level potentials along the line from Eqs.~\eqref{eq:Vqq} to \eqref{eq:Vsys} is equivalent to using the hadron-level contact effective Lagrangians, e.g., for the $\bar{D}^{(\ast)}K^\ast$ systems,
\begin{eqnarray}
\mathcal{L}_{\bar{D}^{(\ast)}K^{\ast}}&=&C_{1}(\bar{\tilde{\mathcal{H}}}\tilde{\mathcal{H}})(K^{\ast\mu\dagger}K_{\mu}^{\ast})
+C_{2}(\bar{\tilde{\mathcal{H}}}\tau_{i}\tilde{\mathcal{H}})(K^{\ast\mu\dagger}\tau_{i}K_{\mu}^{\ast})\nonumber\\
&&+iC_{3}(\bar{\tilde{\mathcal{H}}}\sigma^{\mu\nu}\tilde{\mathcal{H}})(K_{\mu}^{\ast\dagger}K_{\nu}^{\ast})\nonumber\\
&&+iC_{4}(\bar{\tilde{\mathcal{H}}}\sigma^{\mu\nu}\tau_{i}\tilde{\mathcal{H}})(K_{\mu}^{\ast\dagger}\tau_{i}K_{\nu}^{\ast}),
\end{eqnarray}
where $C_i$ ($i=1,\dots,4$) are the LECs, and they can be related to the $c_s$ and $c_a$ in Eq.~\eqref{eq:Vqq}. For example, in our case, $C_1=\frac{1}{2}C_2=-c_s$, $C_3=\frac{1}{2}C_4=-c_a$. The $\tilde{\mathcal{H}}$ denotes the superfield of the $(\bar{D},\bar{D}^\ast)$ doublet, and its expression can be found, e.g., in Ref.~\cite{Meng:2022ozq,Wang:2022jop}. The $K^{\ast}_\mu=(K^{\ast+},K^{\ast0})_\mu^T$ represents the isospin doublet of the $K^\ast$ meson.

In this work, we will stick to the quark-level descriptions, with which we can easily set up the relations between different hadron systems, see the classification in Table~\ref{tab:qc}.

\section{Determining the LECs and calculating the mass spectra}\label{sec:nums}


\subsection{Determining the LECs $c_s$, $c_a$ and $\tilde{c}_s$, $\tilde{c}_a$}\label{sec:lecs}

The hadron-level potentials obtained from Eq.~\eqref{eq:Vsys} will be inserted into the following Lippmann-Schwinger equation (LSE) to search for the bound/virtual state poles,
\begin{eqnarray}\label{eq:intLSE}
t=v+vGt,
\end{eqnarray}
where $t$ is the scattering T-matrix, and $v$ corresponds to the $V_{\text{sys}}$ or $\tilde{V}_{\text{sys}}$ depending on the interacting form of the specified systems. $G$ is the non-relativistic two-body propagator, which reads
\begin{eqnarray}\label{eq:Ge}
  G &=& \int_{0}^{\Lambda}\frac{k^{2}dk}{(2\pi)^{3}}\frac{2\mu}{p^{2}-k^{2}+i\epsilon}\nonumber\\
    &=& \frac{2\mu}{(2\pi)^{3}}\left[p\tanh^{-1}\left(\frac{p}{\Lambda}\right)-\Lambda-\frac{i\pi}{2}p\right],
\end{eqnarray}
where the sharp cutoff regularization is used. The cutoff $\Lambda$ is introduced as a parameter. $p=\sqrt{2\mu(E-m_{\text{th}})}$ is the scattering momentum in which the $E$, $\mu$ and $m_{\text{th}}$ are the center of mass energy, the reduced mass and the threshold of the di-hadron system, respectively.

Because the $v$ is the energy-independent contact potential, so the integral equation~\eqref{eq:intLSE} can be expressed as the algebraic equation,
\begin{eqnarray}\label{eq:invt}
  t^{-1} &=& v^{-1}-G.
\end{eqnarray}
The poles of T-matrix correspond to the zeros of $t^{-1}$. There are two Riemann sheets in the single-channel case. The corresponding $G$ functions for these two sheets respectively read
\begin{eqnarray}
\text{Sheet-I (physical)}&:& G(E+i\epsilon),\nonumber\\
\text{Sheet-II (unphysical)}&:& G(E+i\epsilon)+i\frac{\mu}{4\pi^2} p,
\end{eqnarray}
where $G(E+i\epsilon)$ is the expression given in Eq.~\eqref{eq:Ge}. The analytic continuation of the $G$ function in the second line has been done to reach the unphysical sheet.

There are two independent LECs, $c_s,c_a$ and $\tilde{c}_s,\tilde{c}_a$ in the $V_{qq}$ and $V_{\bar{q}q}$, respectively [see Eqs.~\eqref{eq:Vqq} and \eqref{eq:Vqqbar}]. So we need at least two inputs for each to determine the values of the LECs.
In order to determine the $c_s$ and $c_a$, we will use the three $P_c$ states and the $T_{cc}(3875)$ as inputs. They are regarded as the $\Sigma_c\bar{D}^{(\ast)}$ and $DD^\ast$ bound states, respectively~\cite{Meng:2021jnw,Wang:2019ato,Meng:2019ilv}. Their potentials and the masses with uncertainties are given in Tables~\ref{tab:SigmacDbarsys} and \ref{tab:DDandDbarKsys}, respectively. 
Their masses will correspond to the poles residing at the real-axis below the threshold in the Sheet-I. Inserting the potentials of $P_c$s and $T_{cc}(3875)$ into the LSE~\eqref{eq:invt} we obtain
\begin{eqnarray}
c_{s}&=&146.4\pm10.8\textrm{ GeV}^{-2},\nonumber\\
c_{a}&=&-7.3\pm10.5\textrm{ GeV}^{-2}.
\end{eqnarray}
In the fit, we use a soft value for the cutoff $\Lambda$ in Eq.~\eqref{eq:Ge}, i.e., $\Lambda=0.4$ GeV. The T-matrix should be cutoff-independent, $\partial t/\partial\Lambda=0$,  which can be achieved via using a cutoff-dependent potential $v$ that satisfies $\partial v^{-1}/\partial \Lambda=\partial G/\partial \Lambda$. However, if the potentials are derived basing on the sprit of effective field theory, the $\Lambda$ is restricted in a range instead of reaching arbitrarily large values in the renormalization group flow~\cite{Machleidt:2011zz}. In Sec.~\ref{sec:intro}, we have analyzed that the typical hard scale is $m^{\text{QM}}_s\sim 500$ MeV, so the reasonable choice for $\Lambda$ is $\Lambda\lesssim m_s^{\text{QM}}$. We will use the $\Lambda=0.4$ GeV throughout this work.

\begin{table*}[hptb]
\renewcommand{\arraystretch}{1.5}
\caption{The $I(J^P)$ quantum numbers, effective potentials, and possible molecule candidates in the $\Sigma_c\bar{D}^{(\ast)}$ systems. The $E_B$ denotes the mass (in units of MeV) of the corresponding $P_c$ as a bound state. The masses with errors are taken from the LHCb data~\cite{LHCb:2019kea}. \label{tab:SigmacDbarsys}}
\setlength{\tabcolsep}{3.9mm}
{
\centering
\begin{tabular}{ccccccccc}
\toprule[0.8pt]
Systems & $I(J^{P})$ & $\langle\mathcal{O}_{1}\rangle$ & $\langle\mathcal{O}_{2}\rangle$ & $\langle\mathcal{O}_{3}\rangle$ & $\langle\mathcal{O}_{4}\rangle$ & $V_{\textrm{sys}}$ & $E_{B}$ & States\tabularnewline
\hline
$\Sigma_{c}\bar{D}$ & $\frac{1}{2}\left(\frac{1}{2}^{-}\right)$ & $1$ & $-4$ & $0$ & $0$ & $-3c_{s}$ & $4311.9\pm0.7_{-0.6}^{+6.8}$ & $P_{c}(4312)$ {[}Input{]}\tabularnewline
\hline
\multirow{2}{*}{$\Sigma_{c}\bar{D}^{\ast}$} & $\frac{1}{2}\left(\frac{1}{2}^{-}\right)$ & $1$ & $-4$ & $-\frac{4}{3}$ & $\frac{16}{3}$ & $-3c_{s}+4c_{a}$ & $4440.3\pm1.3_{-4.7}^{+4.1}$ & $P_{c}(4440)$ {[}Input{]}\tabularnewline
 & $\frac{1}{2}\left(\frac{3}{2}^{-}\right)$ & $1$ & $-4$ & $\frac{2}{3}$ & $-\frac{8}{3}$ & $-3c_{s}-2c_{a}$ & $4457.3\pm0.6_{-1.7}^{+4.1}$ & $P_{c}(4457)$ {[}Input{]}\tabularnewline
\bottomrule[0.8pt]
\end{tabular}
}
\end{table*}

\begin{figure}[hptb]
\begin{centering}
    \scalebox{0.8}{\includegraphics[width=\linewidth]{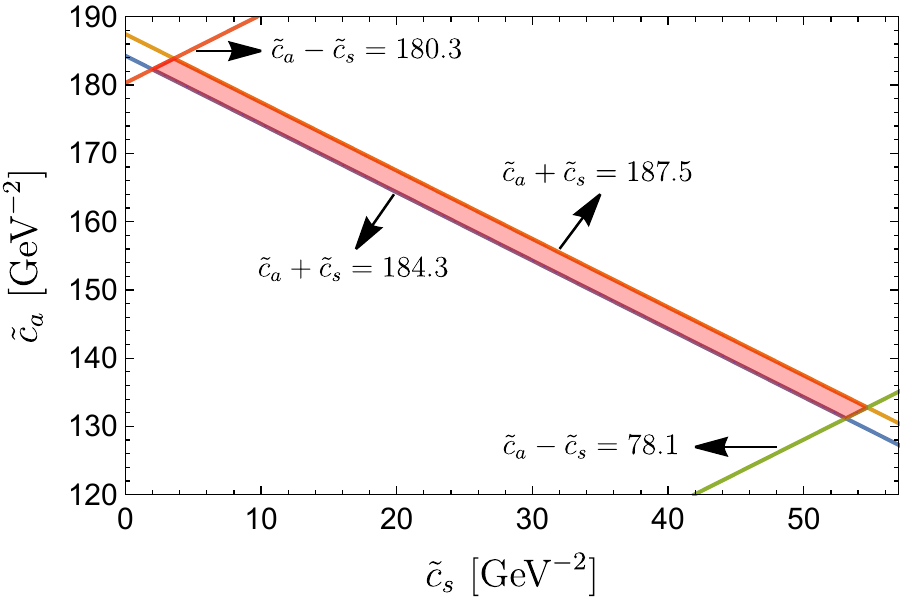}}
    \caption{The constrained region (denoted as the red rectangle area) for the LECs $\tilde{c}_s$ and $\tilde{c}_a$ when using the $X(3872)$ and $Z_c(3900)$ as the inputs.\label{fig:paras_range}}
\end{centering}
\end{figure}

The $\tilde{c}_s$ and $\tilde{c}_a$ will be determined using the $X(3872)$ and $Z_c(3900)$ as inputs. The $X(3872)$ and $Z_c(3900)$ are still very disputed although they have been observed for many years:
\begin{itemize}
  \item The multifaceted nature of the $X(3872)$ has been subject to extensive debate, with interpretations ranging from a $D\bar{D}^\ast$ bound state to a conventional charmonium $\chi_{c1}(2P)$ to a hybrid model involving a mixture of $D\bar{D}^\ast$ with the $c\bar{c}$ core~\cite{Chen:2016qju,Guo:2017jvc,Liu:2019zoy,Meng:2021kmi,Meng:2022ozq,Wang:2023ovj}.  While Esposito \etal's analysis of prompt productions~\cite{CMS:2013fpt,ALICE:2015oer,ALICE:2015wav} suggested the presence of a compact $c\bar{c}$ core at short distances~\cite{Esposito:2015fsa}, a compelling body of experimental evidence has emerged in favor of a predominantly molecular character for the $X(3872)$. For example, the line-shape analysis of $X(3872)$ by the LHCb Collaboration revealed that the probability to find a compact component in the $X(3872)$ wave function is less than a third~\cite{LHCb:2020xds}. Within a analysis~\cite{Baru:2021ldu} using the LHCb data but with an improved formalism, it was concluded that the component of the hadronic molecule is over $90\%$. Recently, the line-shape analysis of $X(3872)$ by the BESIII Collaborations yields the $82\%$ compositeness~\cite{BESIII:2023hml}. Moreover, theoretical studies within diverse frameworks have corroborated this molecular dominance, with the compact component estimated to be exceptionally small~\cite{Ortega:2009hj,Wang:2023ovj}. 
  Regarding the isospin breaking decays, Gamermann \etal~have demonstrated that the couplings of the $X(3872)$ to charged and neutral $D$ mesons exhibit striking similarity, with at most $1.4\%$ of isospin violation~\cite{Gamermann:2009fv}. The isospin violating decay is attributed to the interplay of the sizable width of the intermediate $\rho$ meson and the amplification effects within the two-body propagators~\cite{Gamermann:2009fv}.
  In light of the above reasonings, and given our focus on long-range dynamics, we will adopt the approximation of the $X(3872)$ as an isoscalar $D\bar{D}^\ast$ molecule for the purposes of this investigation.
  \item The enigmatic nature of the $Z_c(3900)$ has sparked intriguing theoretical debates, with its characterization as either a virtual state or a resonance intricately linked to the energy dependence of the underlying interaction potential~\cite{Albaladejo:2015lob,Wang:2020dko,Cheng:2023vyv}. In studies employing energy-independent potentials, the $Z_c(3900)$ emerges as a virtual state, whereas energy-dependent potentials often favor a resonant interpretation. This dichotomy is strikingly illustrated by the work of Albaladejo \etal~\cite{Albaladejo:2015lob}, who performed simultaneous fits to the $J/\psi\pi$ and $D\bar{D}^\ast$ invariant mass distributions using both energy-independent and energy-dependent potentials. Notably, both approaches yielded remarkably similar quality in describing the experimental data, yet resulted in virtual state and resonance solutions, respectively. We also investigated the line-shapes of $Z_c(3900)$ [as well as the $Z_c(4020)$ and $Z_b(10610)$, $Z_b(10650)$] in the open-charm (bottom) channels within the chiral effective field theory~\cite{Wang:2020dko}, and we noticed that the data can be fairly described in the single-channel case. Further insights into this model-dependent duality are provided by the comprehensive analyses of Chen \etal~\cite{Chen:2023def}. In the present investigation, we adhere to the framework of constant potentials, thus positioning the $Z_c(3900)$ as an isovector virtual state originating from $D\bar{D}^\ast$ interactions.
\end{itemize}
  The mass of $X(3872)$ is set to be below the $D^0\bar{D}^{\ast0}$ threshold about $[50,500]$ keV, while the pole mass of $Z_c(3900)$ is set to be below the $D\bar{D}^\ast$ threshold about $[1,50]$ MeV at the real-axis in Sheet-II. Their potentials and (pole) masses are listed in Table~\ref{tab:DDbarandDKsys}. With these two inputs we obtain
\begin{eqnarray}
184.3\textrm{ GeV}^{-2}<&\tilde{c}_{a}+\tilde{c}_{s}&<187.5\textrm{ GeV}^{-2},\nonumber\\
78.1\textrm{ GeV}^{-2}<&\tilde{c}_{a}-\tilde{c}_{s}&<180.3\textrm{ GeV}^{-2}.
\end{eqnarray}
One can notice that the $\tilde{c}_{s}$ and $\tilde{c}_{a}$ are confined in a rectangle region, which is plotted in Fig.~\ref{fig:paras_range}.

\subsection{Molecule spectra of the $D^{(\ast)}D^{(\ast)}$, $\bar{D}^{(\ast)}K^\ast$ and $D^{(\ast)}\bar{D}^{(\ast)}$, $D^{(\ast)}K^\ast$ systems}\label{sec:spec}


\subsubsection*{$D^{(\ast)}D^{(\ast)}$ and $\bar{D}^{(\ast)}K^\ast$ systems}

The numerical results for the $D^{(\ast)}D^{(\ast)}$ and $\bar{D}^{(\ast)}K^\ast$ systems are collected in Table~\ref{tab:DDandDbarKsys}. One can see that apart from the $T_{cc}(3875)$, there also exists a bound state $T_{cc}(4015)$ in the $D^\ast D^\ast$ system with the same quantum numbers as the $T_{cc}(3875)$, whereas no bound or virtual states can be found in the isovector channels (the similar behavior was also reported in the double-bottom $\bar{B}^{\ast}\bar{B}^{\ast}$ systems~\cite{Wang:2018atz}). The possible molecular states in the $D^\ast D^{(\ast)}$ systems have been studied in many works~\cite{Li:2012ss,Xu:2017tsr,Liu:2019stu,Ding:2020dio,Dong:2021bvy,Chen:2021cfl}.

\begin{figure}[htbp]
\begin{centering}
    \scalebox{0.8}{\includegraphics[width=\linewidth]{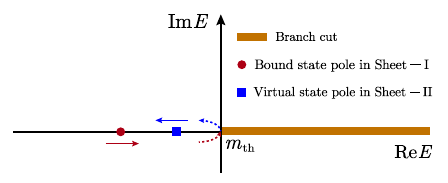}}
    \caption{The pole trajectory with decreasing the strength of the attractive potential. The solid arrows denote the directions of the movement of poles when the interaction becomes weaker, while the dashed arrow means that the pole will move on to the Sheet-II when the attractive potential reaches a small critical value.\label{fig:poletraj}}
\end{centering}
\end{figure}

As analyzed in Sec.~\ref{sec:intro}, the interactions of the $\bar{D}^{(\ast)} K^\ast$ systems can be related to those of $D^{(\ast)}D^{(\ast)}$, e.g., see the correspondence in Eq.~\eqref{eq:corresp}. From Table~\ref{tab:DDandDbarKsys}, one can see that the potential of $\bar{D}^\ast K^\ast$ equals to that of $D^\ast D^{(\ast)}$ in the $0(1^+)$ channel, while the $\bar{D}^\ast K^\ast$ system in the $0(1^+)$ channel becomes a virtual state rather than the bound state as that of the $D^\ast D^{(\ast)}$. The reason is the reduced mass of $\bar{D}^\ast K^\ast$ ($\sim0.6$ GeV) is smaller than that of $D^\ast D^{(\ast)}$ ($\sim1$ GeV). We know that the emergence of bound states results from the attractive potential overwhelms the kinetic energy, recalling that the term $\bm{p}^2/(2\mu)+V(\bm r)$ in the Schr\"odinger equation. Thus, the effective attraction will be weakened with the decreasing of the reduced mass $\mu$. With the terminology of our framework, it amounts to saying that the bound state pole in Sheet-I will move on to the Sheet-II (virtual state pole) when the attractive potential reaches a small critical value, e.g., see the vivid illustration in Fig.~\ref{fig:poletraj}. One will see in the following parts that the whole $\bar{D}^{(\ast)} K^\ast$ and $D^{(\ast)} K^\ast$ systems with specific quantum numbers correspond to virtual states.

In the isoscalar $\bar{D}K^\ast$ and $\bar{D}^\ast K^\ast$ systems, a total of four virtual states have been identified. These are denoted as $\bar{D}K^\ast[0(1^+)]$ and $\bar{D}^\ast K^\ast[0(0^+),0(1^+),0(2^+)]$, respectively. These pole positions are found to be close to the respective thresholds. However, much like the $D^{(\ast)}D^{(\ast)}$ systems, no bound or virtual states have been observed in the isovector channels. The scalar-isoscalar state within the $\bar{D}^\ast K^\ast$ system is presumed to correspond to the LHCb-observed $T_{cs0}(2900)$ in the $D^-K^+$ channel~\cite{LHCb:2020bls, LHCb:2020pxc}. The $0(0^+)$ state can easily decay into the $\bar{D}K$ channel through S-wave transitions, for instance, via the pion exchange. However, for the $0(1^+)$ states in the $\bar{D}K^\ast$ and $\bar{D}^\ast K^\ast$ systems, transitions to $\bar{D}K$ are forbidden due to parity constraints. Thus, the most probable channel for detecting these two states is $\bar{D}^\ast K$. For instance, the vector-isoscalar $\bar{D}K^\ast$ and $\bar{D}^\ast K^\ast$ states can readily decay into $\bar{D}^\ast K$ through pion exchange. Consequently, it is expected that there should be a peak at the $\bar{D} K^\ast$ ($2.76$ GeV) and $\bar{D}^\ast K^\ast$ ($2.9$ GeV) thresholds in the $\bar{D}^\ast K$ channel, respectively. As for the $0(2^+)$ state, it can decay into the $\bar{D}K$ channel through D-wave transitions. Therefore, it is highly plausible that the invariant mass distribution of $D^-K^+$~\cite{LHCb:2020bls, LHCb:2020pxc} encompasses the contributions from the $0(2^+)$ state. The vector and tensor partners of the $T_{cs0}(2900)$ have been suggested for exploration in $B$ decay channels~\cite{Dai:2022qwh, Dai:2022htx, Bayar:2022wbx}.

\begin{table*}[!ht]
\renewcommand{\arraystretch}{1.4}
\caption{The $I(J^{P})$ quantum numbers, effective potentials, and possible molecule candidates in the $D^{(\ast)}D^{(\ast)}$ and $\bar{D}^{(\ast)}K^\ast$ systems. The $E_B/E_V$ denotes the (pole) mass (in units of MeV) of the corresponding state as a bound/virtual state. The predicted state is marked with bold type, while the symbol `$-$' means that no bound/virtual states can be found in the corresponding systems. The mass of $T_{cc}(3875)$ is taken from the LHCb data~\cite{LHCb:2021auc}. \label{tab:DDandDbarKsys}}
\setlength{\tabcolsep}{3.8mm}
{
\centering
\begin{tabular}{ccccccccc}
\toprule[0.8pt]
Systems & $I(J^{P})$ & $\langle\mathcal{O}_{1}\rangle$ & $\langle\mathcal{O}_{2}\rangle$ & $\langle\mathcal{O}_{3}\rangle$ & $\langle\mathcal{O}_{4}\rangle$ & $V_{\textrm{sys}}$ & $E_{B}/E_V$ & States\tabularnewline
\hline
$DD$ & $1(0^{+})$ & $\frac{1}{2}$ & $1$ & $0$ & $0$ & $\frac{3}{2}c_{s}$ & $-$ & $-$\tabularnewline
\hline
\multirow{2}{*}{$DD^{\ast}$} & $0(1^{+})$ & $\frac{1}{2}$ & $-3$ & $-\frac{1}{2}$ & $3$ & $-\frac{5}{2}c_{s}+\frac{5}{2}c_{a}$ & $[3874.74\pm0.04]_B$ & $T_{cc}(3875)$ {[}Input{]}\tabularnewline
 & $1(1^{+})$ & $\frac{1}{2}$ & $1$ & $\frac{1}{2}$ & $1$ & $\frac{3}{2}c_{s}+\frac{3}{2}c_{a}$ & $-$ & $-$\tabularnewline
\hline
\multirow{3}{*}{$D^{\ast}D^{\ast}$} & $0(1^{+})$ & $\frac{1}{2}$ & $-3$ & $-\frac{1}{2}$ & $3$ & $-\frac{5}{2}c_{s}+\frac{5}{2}c_{a}$ & $\left[4015.6_{-2.3}^{+0.8}\right]_B$ & $\bm{T_{cc}(4015)}$\tabularnewline
 & $1(0^{+})$ & $\frac{1}{2}$ & $1$ & $-1$ & $-2$ & $\frac{3}{2}c_{s}-3c_{a}$ & $-$ & $-$\tabularnewline
 & $1(2^{+})$ & $\frac{1}{2}$ & $1$ & $\frac{1}{2}$ & $1$ & $\frac{3}{2}c_{s}+\frac{3}{2}c_{a}$ & $-$ & $-$\tabularnewline
\hline
\multirow{2}{*}{$\bar{D}K^{\ast}$} & $0(1^{+})$ & $\frac{1}{2}$ & $-3$ & $0$ & $0$ & $-\frac{5}{2}c_{s}$ & $\left[2752.5_{-3.6}^{+2.6}\right]_{V}$ & $\bm{T_{cs1}^{f}(2760)}$\tabularnewline
 & $1(1^{+})$ & $\frac{1}{2}$ & $1$ & $0$ & $0$ & $\frac{3}{2}c_{s}$ & $-$ & $-$\tabularnewline
\hline
\multirow{6}{*}{$\bar{D}^{\ast}K^{\ast}$} & $0(0^{+})$ & $\frac{1}{2}$ & $-3$ & $-1$ & $6$ & $-\frac{5}{2}c_{s}+5c_{a}$ & $\left[2897.8_{-9.1}^{+2.6}\right]_{V}$ & $T_{cs0}^f(2900)$\tabularnewline
 & $0(1^{+})$ & $\frac{1}{2}$ & $-3$ & $-\frac{1}{2}$ & $3$ & $-\frac{5}{2}c_{s}+\frac{5}{2}c_{a}$ & $\left[2896.6_{-6.3}^{+3.0}\right]_{V}$ & $\bm{T_{cs1}^{f}(2900)}$\tabularnewline
 & $0(2^{+})$ & $\frac{1}{2}$ & $-3$ & $\frac{1}{2}$ & $-3$ & $-\frac{5}{2}c_{s}-\frac{5}{2}c_{a}$ & $\left[2892.8_{-9.8}^{+5.0}\right]_{V}$ & $\bm{T_{cs2}^{f}(2900)}$\tabularnewline
 & $1(0^{+})$ & $\frac{1}{2}$ & $1$ & $-1$ & $-2$ & $\frac{3}{2}c_{s}-3c_{a}$ & $-$ & $-$\tabularnewline
 & $1(1^{+})$ & $\frac{1}{2}$ & $1$ & $-\frac{1}{2}$ & $-1$ & $\frac{3}{2}c_{s}-\frac{3}{2}c_{a}$ & $-$ & $-$\tabularnewline
 & $1(2^{+})$ & $\frac{1}{2}$ & $1$ & $\frac{1}{2}$ & $1$ & $\frac{3}{2}c_{s}+\frac{3}{2}c_{a}$ & $-$ & $-$\tabularnewline
\bottomrule[0.8pt]
\end{tabular}
}
\end{table*}

\subsubsection*{$D^{(\ast)}\bar{D}^{(\ast)}$ and $D^{(\ast)}K^\ast$ systems}

The numerical results for the $D^{(\ast)}\bar{D}^{(\ast)}$ and $D^{(\ast)}K^\ast$ systems are provided in Table~\ref{tab:DDbarandDKsys}, with the $X(3872)$ and $Z_c(3900)$ serving as our inputs. It is evident that the spectrum of $D^{(\ast)}\bar{D}^{(\ast)}$ is richer compared to that of $D^{(\ast)}D^{(\ast)}$ as shown in Table~\ref{tab:DDandDbarKsys}. This is primarily due to the greater diversity of quantum numbers in $D^{(\ast)}\bar{D}^{(\ast)}$, as there are no symmetry constraints on their wave functions. However, it is important to note that not all channels within $D^{(\ast)}\bar{D}^{(\ast)}$ have bound or virtual states. For example, the states with opposite $C(G)$ parity to those of $X(3872)$ and $Z_c(3900)$ have been proven to be nonexistent. The $C$ parity odd partner of $X(3872)$ has been searched for by the Babar~\cite{BaBar:2004iez}, Belle~\cite{Belle:2013vio} and LHCb~\cite{LHCb:2022oqs} Collaborations, but no signal was observed near the $D\bar{D}^\ast$ threshold in the $J/\psi\eta$ invariant mass spectrum.

One can see that the $0^+(2^{++})$ and $1^+(1^{+-})$ states in the $D^\ast\bar{D}^\ast$ systems exist as the heavy quark spin symmetry partners of $X(3872)$ and $Z_c(3900)$, respectively. The former was predicted long ago by T\"ornqvist~\cite{Tornqvist:1993ng} with a mass $4015$ MeV, and it received intensive studies after the observation of $X(3872)$~\cite{Nieves:2012tt,Hidalgo-Duque:2012rqv,Sun:2012zzd,Guo:2013sya,Albaladejo:2015dsa,Baru:2016iwj}. Recently, the BESIII Collaboration searched for the $0^+(2^{++})$ state in the $D\bar{D}$ invariant mass spectrum by the process $e^+e^-\to\pi^+\pi^-D\bar{D}$~\cite{BESIII:2019tdo}, but no obvious signal was found. Very recently, the Belle Collaboration studied the $\gamma\gamma\to\gamma\psi(2S)$~\cite{Belle:2021nuv}, and reported a hint of new state (with global significance $2.8\sigma$) with mass and width $[m,\Gamma]=[4014.3\pm4.0\pm1.5,4\pm11\pm6]$ MeV, which was conjectured to be the tensor-isoscalar partner of $X(3872)$~\cite{Duan:2022upr,Yue:2022gym,Shi:2023mer}. The $1^+(1^{+-})$ state in $D^\ast\bar{D}^\ast$ system shall correspond to the known $Z_c(4020)$~\cite{BESIII:2013ouc}.

It is worth noting that we have also identified a bound state with quantum numbers $1^-(0^{++})$ in the $D^\ast\bar{D}^\ast$ system, with a mass of approximately $4010$ MeV. This state can decay through open-charm and hidden-charm channels, including the $D\bar{D}$ and $\eta_c\pi$, $J/\psi\rho$, and $\chi_{c1}\pi$ (via P-wave) modes. Among these, the $\eta_c\pi$ channel has the highest phase space. It is important to mention that the LHCb Collaboration recently reported evidence (with a significance of more than three standard deviations) of an exotic resonance denoted as $Z_c(4100)^-$ in the $\eta_c\pi^-$ invariant mass distribution from the decay $B^0\to\eta_c\pi^-K^+$~\cite{LHCb:2018oeg}. The measured mass and width for this state are $[m,\Gamma]=[4096\pm20^{+18}_{-22},152\pm58^{+60}_{-35}]$ MeV. The spin-parity assignment is not definitively determined within the present statistics, with possibilities including $J^P=0^+$ and $1^-$. Consequently, further data is required to ascertain whether the $1^-(0^{++})$ state predicted in our work and the $Z_c(4100)^-$ observed in Ref.~\cite{LHCb:2018oeg} are indeed the same entity.

Another intriguing channel is the $D\bar{D}$ with $0^+(0^{++})$ quantum numbers, where the potential depends solely on $\tilde{c}_s$. From Fig.~\ref{fig:paras_range}, it is apparent that $\tilde{c}_s$ spans a range from $2$ to $55$ GeV$^{-2}$. For large values of $\tilde{c}_s$, a virtual state emerges in this channel. However, as $\tilde{c}_s$ decreases (as illustrated in Fig.~\ref{fig:poletraj}), the virtual state pole moves away from the $D\bar{D}$ threshold, eventually vanishing when $\tilde{c}_s$ falls below a critical value. Therefore, in Table~\ref{tab:DDbarandDKsys}, we use the notation $[\sim x]_V$ for this channel, where $x$ represents the pole position when $\tilde{c}_s$ reaches its maximum value within the confined region shown in Fig.~\ref{fig:paras_range}. The potential existence of a molecular state in the scalar-isoscalar $D\bar{D}$ system has been investigated in previous studies (Refs.~\cite{Zhang:2006ix,Gamermann:2006nm,Hidalgo-Duque:2012rqv,Nieves:2012tt,Prelovsek:2020eiw}). Various processes, such as $\psi(3770)\to\gamma\eta\eta^\prime$~\cite{Xiao:2012iq}, $B^{0,+}\to D^0\bar{D}^0K^{0,+}$~\cite{Dai:2015bcc}, $\psi(3770)\to\gamma D^0\bar{D}^0$~\cite{Dai:2020yfu}, and $\Lambda_b\to\Lambda D\bar{D}$~\cite{Wei:2021usz}, have been proposed as potential avenues for the search of this state. Notably, the BESIII Collaboration recently conducted searches for this state in the decays $\psi(3770)\to\gamma\eta\eta^\prime,\gamma\pi^+\pi^-J/\psi$~\cite{BESIII:2023bgk}, but no significant signals were observed. In our calculations, the existence of the virtual state solution is contingent on the value of $\tilde{c}_s$. If $\tilde{c}_s$ is fine-tuned to approach zero [equivalent to using a pole mass very close to threshold for the $Z_c(3900)$], this state will no longer persist. This theoretically proposed state comprises the two lightest charm mesons, resulting in a limited range of decay modes. The dominant decay channel has been suggested to be $\eta\eta^\prime$ via $D\bar{D}$ annihilation~\cite{Xiao:2012iq}. The restricted and narrow-width decay modes make it challenging to directly measure this state in experiments. However, as we will see in the following sections, the $0(1^+)$ channel in the $DK^\ast$ system can play a crucial role in indirectly inferring the existence of this state.

We then move onto the $D^{(\ast)}K^\ast$ systems. It is worth noting that the $0(1^+)$ $DK^\ast$ system shares the same potential as the $0^+(0^{++})$ $D\bar{D}$ system. Consequently, virtual state solutions also exist in the $0(1^+)$ $DK^\ast$ system. However, the pole position is considerably far from the $DK^\ast$ threshold due to its relatively smaller reduced mass. Since $DK^\ast$ with $0(1^+)$ quantum numbers can decay into $D^\ast K$ via one-pion exchange, experimental searches for this state can be conducted in the $D^\ast K$ channel. If this state indeed exists as a virtual state, a peak would emerge at the $DK^\ast$ threshold ($2.76$ GeV) in the $D^\ast K$ invariant mass spectrum. The existence of a $0(1^+)$ $DK^\ast$ molecule would also imply the presence of a $0^+(0^{++})$ $D\bar{D}$ molecule. This is why we suggest inferring the existence of the $0^+(0^{++})$ $D\bar{D}$ state from the $0(1^+)$ $DK^\ast$ system.

In the $D^\ast K^\ast$ system, there exists a tensor-isoscalar partner of the $X(3872)$, corresponding to a virtual state with the pole located very close to the $D^\ast K^\ast$ threshold. However, its $0(0^+)$ and $0(1^+)$ partners have been shown to be nonexistent. This state can be searched for in the $DK$ channel, with additional angle distribution analysis for the $DK$ system. For instance, the angle distribution of the $DK$ system should exhibit evident D-wave behavior, as the $0(2^+)$ $D^\ast K^\ast$ decays into $DK$ via the D-wave transition. Moreover, experimental efforts to find this state can also be used to indirectly infer the existence of the $0^+(2^{++})$ partner of the $X(3872)$ in the $D^\ast\bar{D}^\ast$ system.

In the isovector $D^\ast K^\ast$ system, there are two charmed strange partners of the $Z_c(3900)$:
(i) The first one has quantum numbers $1(0^+)$, corresponding to the  $T_{c\bar{s}0}^a(2900)$ observed by LHCb~\cite{LHCb:2022sfr,LHCb:2022lzp}. It is a virtual state with the pole close to the $D^\ast K^\ast$ threshold. Its primary decay modes are expected to be $D_s\pi$ and $DK$. Interestingly, this state is a counterpart of the $1^-(0^{++})$ bound state predicted in the $D^\ast\bar{D}^\ast$ system, i.e., the $Z_c(4010)$. Therefore, the observed channel $D_s\pi$ of $T_{c\bar{s}0}^a(2900)$ corresponds to the reported channel $\eta_c\pi$ of the $Z_c(4010)$. It would be enlightening to measure the partial width ratio of $T_{c\bar{s}0}^a(2900)$ in experiments, denoted as $\mathcal{R}{c\bar{s}}=\Gamma[T_{c\bar{s}0}^a(2900)\to DK]/\Gamma[T_{c\bar{s}0}^a(2900)\to D_s\pi]$. The analogous quantity for the $Z_c(3900)$ is $\mathcal{R}{c\bar{c}}=\Gamma[Z_c(3900)\to D\bar{D}^\ast]/\Gamma[Z_c(3900)\to J/\psi\pi]=6.2\pm1.1\pm2.7$~\cite{BESIII:2013qmu}. Consequently, we conservatively estimate $\mathcal{R}{c\bar{s}}\sim10$, considering the ample phase space available for the $DK$ channel. We note that the calculation in Ref.~\cite{Yue:2022mnf} is consistent with our estimation. (ii) The second one has quantum numbers $1(1^+)$ and it is the counterpart of $Z_c(3900)$. Its pole mass varies from tens to a few hundred MeV below the $D^\ast K^\ast$ threshold. The ideal decay channels for this state are $D_s^\ast\pi$ and $D^\ast K$. In particular, the $D_s^\ast\pi$ channel is analogous to the $J/\psi\pi$ decay mode of the $Z_c(3900)$. Thus, we roughly estimate the partial width ratio $\Gamma[T_{c\bar{s}1}^a(2900)\to D^\ast K]/\Gamma[T_{c\bar{s}1}^a(2900)\to D_s^\ast\pi]\sim10$. Experiments can search for this state either in the $D_s^\ast\pi$ or $D^\ast K$ invariant mass spectrum. This concludes the discussion on the charmed strange partners of $Z_c(3900)$ in the isovector $D^\ast K^\ast$ system.

\begin{table*}[!ht]
\renewcommand{\arraystretch}{1.4}
\caption{The $I^{(G)}(J^{P(C)})$ quantum numbers, effective potentials, and possible molecule candidates in the $D^{(\ast)}\bar{D}^{(\ast)}$ and $D^{(\ast)} K^\ast$ systems. The notations are the same as those in Table~\ref{tab:DDandDbarKsys}, while the superscript $\sharp$ means that this state will vanish when the attraction is too weak.\label{tab:DDbarandDKsys}}
\setlength{\tabcolsep}{3.4mm}
{
\centering
\begin{tabular}{ccccccccc}
\toprule[0.8pt]
Systems  & $I^{(G)}(J^{P(C)})$ & $\langle\mathcal{O}_{1}\rangle$ & $\langle\mathcal{O}_{2}\rangle$ & $\langle\mathcal{O}_{3}\rangle$ & $\langle\mathcal{O}_{4}\rangle$ & $\tilde{V}_{\textrm{sys}}$ & $E_{B}/E_{V}$ & States\tabularnewline
\hline
\multirow{2}{*}{$D\bar{D}$} & $0^{+}(0^{++})$ & $\frac{1}{2}$ & $-3$ & $0$ & $0$ & $-\frac{5}{2}\tilde{c}_{s}$ & $\left[\sim3696.5\right]_{V}$ & $\bm{X(3700)}^{\sharp}$\tabularnewline
 & $1^{-}(0^{++})$ & $\frac{1}{2}$ & $1$ & $0$ & $0$ & $\frac{3}{2}\tilde{c}_{s}$ & $-$ & $-$\tabularnewline
\hline
\multirow{4}{*}{$D\bar{D}^{\ast}$} & $0^{+}(1^{++})$ & $\frac{1}{2}$ & $-3$ & $\frac{1}{2}$ & $-3$ & $-\frac{5}{2}\tilde{c}_{s}$$-\frac{5}{2}\tilde{c}_{a}$ & $\left[3871.2,3871.6\right]_{B}$ & $X(3872)$ {[}Input{]}\tabularnewline
 & $0^{-}(1^{+-})$ & $\frac{1}{2}$ & $-3$ & $-\frac{1}{2}$ & $3$ & $-\frac{5}{2}\tilde{c}_{s}$$+\frac{5}{2}\tilde{c}_{a}$ & $-$ & $-$\tabularnewline
 & $1^{+}(1^{+-})$ & $\frac{1}{2}$ & $1$ & $-\frac{1}{2}$ & $-1$ & $\frac{3}{2}\tilde{c}_{s}-\frac{3}{2}\tilde{c}_{a}$ & $\left[3825.8,3874.8\right]_{V}$ & $Z_{c}(3900)$ {[}Input{]}\tabularnewline
 & $1^{-}(1^{++})$ & $\frac{1}{2}$ & $1$ & $\frac{1}{2}$ & $1$ & $\frac{3}{2}\tilde{c}_{s}+\frac{3}{2}\tilde{c}_{a}$ & $-$ & $-$\tabularnewline
\hline
\multirow{6}{*}{$D^{\ast}\bar{D}^{\ast}$} & $0^{+}(0^{++})$ & $\frac{1}{2}$ & $-3$ & $-1$ & $6$ & $-\frac{5}{2}\tilde{c}_{s}+5\tilde{c}_{a}$ & $-$ & $-$\tabularnewline
 & $0^{-}(1^{+-})$ & $\frac{1}{2}$ & $-3$ & $-\frac{1}{2}$ & $3$ & $-\frac{5}{2}\tilde{c}_{s}+\frac{5}{2}\tilde{c}_{a}$ & $-$ & $-$\tabularnewline
 & $0^{+}(2^{++})$ & $\frac{1}{2}$ & $-3$ & $\frac{1}{2}$ & $-3$ & $-\frac{5}{2}\tilde{c}_{s}$$-\frac{5}{2}\tilde{c}_{a}$ & $\left[4011.8,4012.2\right]_{B}$ & $\bm{X(4012)}$\tabularnewline
 & $1^{-}(0^{++})$ & $\frac{1}{2}$ & $1$ & $-1$ & $-2$ & $\frac{3}{2}\tilde{c}_{s}-3\tilde{c}_{a}$ & $\left[4007.2,4016.7\right]_{B}$ & $\bm{Z_c(4010)}$\tabularnewline
 & $1^{+}(1^{+-})$ & $\frac{1}{2}$ & $1$ & $-\frac{1}{2}$ & $-1$ & $\frac{3}{2}\tilde{c}_{s}-\frac{3}{2}\tilde{c}_{a}$ & $\left[3973.6,4014.5\right]_{V}$ & $Z_{c}(4020)$\tabularnewline
 & $1^{-}(2^{++})$ & $\frac{1}{2}$ & $1$ & $\frac{1}{2}$ & $1$ & $\frac{3}{2}\tilde{c}_{s}+\frac{3}{2}\tilde{c}_{a}$ & $-$ & $-$\tabularnewline
\hline
\multirow{2}{*}{$DK^{\ast}$} & $0(1^{+})$ & $\frac{1}{2}$ & $-3$ & $0$ & $0$ & $-\frac{5}{2}\tilde{c}_{s}$ & $\left[\sim2586.0\right]_{V}$ & $\bm{T_{c\bar{s}0}^{f}(2760)}^{\sharp}$\tabularnewline
 & $1(1^{+})$ & $\frac{1}{2}$ & $1$ & $0$ & $0$ & $\frac{3}{2}\tilde{c}_{s}$ & $-$ & $-$\tabularnewline
\hline
\multirow{6}{*}{$D^{\ast}K^{\ast}$} & $0(0^{+})$ & $\frac{1}{2}$ & $-3$ & $-1$ & $6$ & $-\frac{5}{2}\tilde{c}_{s}+5\tilde{c}_{a}$ & $-$ & $-$\tabularnewline
 & $0(1^{+})$ & $\frac{1}{2}$ & $-3$ & $-\frac{1}{2}$ & $3$ & $-\frac{5}{2}\tilde{c}_{s}+\frac{5}{2}\tilde{c}_{a}$ & $-$ & $-$\tabularnewline
 & $0(2^{+})$ & $\frac{1}{2}$ & $-3$ & $\frac{1}{2}$ & $-3$ & $-\frac{5}{2}\tilde{c}_{s}-\frac{5}{2}\tilde{c}_{a}$ & $\left[2900.2,2900.3\right]_{V}$ & $\bm{T_{c\bar{s}2}^{f}(2900)}$\tabularnewline
 & $1(0^{+})$ & $\frac{1}{2}$ & $1$ & $-1$ & $-2$ & $\frac{3}{2}\tilde{c}_{s}-3\tilde{c}_{a}$ & $\left[2887.6,2900.5\right]_{V}$ & $T_{c\bar{s}0}^{a}(2900)$\tabularnewline
 & $1(1^{+})$ & $\frac{1}{2}$ & $1$ & $-\frac{1}{2}$ & $-1$ & $\frac{3}{2}\tilde{c}_{s}-\frac{3}{2}\tilde{c}_{a}$ & $\left[2676.2,2876.3\right]_{V}$ & $\bm{T_{c\bar{s}1}^{a}(2900)}$\tabularnewline
 & $1(2^{+})$ & $\frac{1}{2}$ & $1$ & $\frac{1}{2}$ & $1$ & $\frac{3}{2}\tilde{c}_{s}+\frac{3}{2}\tilde{c}_{a}$ & $-$ & $-$\tabularnewline
\bottomrule[0.8pt]
\end{tabular}
}
\end{table*}

\section{Discussions on the line-shapes of $T_{c\bar{s}0}^a(2900)$}\label{sec:linesh}

From Table~\ref{tab:DDbarandDKsys}, we see that the recently observed $T_{c\bar{s}0}^a(2900)$ can be well identified as a near-threshold virtual state. In contrast to the truly bound state which behaves as a real particle, the virtual state is usually called the anti-bound state since its wave function is not localized~\cite{Newton:1982qc}. But the virtual state can indeed induce observable effects, such as in the nuclear reactions~\cite{Mukhamedzhanov:2009de,Konobeevski:2016lsq} and the experimental line-shapes~\cite{Frazer:1964zz,Badalian:1981xj}, etc. One can also find a new method to probe the virtual state pole using improved complex scaling method in Ref.~\cite{Chen:2023eri}. In the following, we will explore the influence of the virtual state $T_{c\bar{s}0}^a(2900)$ on the $D_s\pi$ invariant mass spectrum. To this end, we introduce a virtual state pole within the coupled-channel framework. We consider a two-channel coupling for $T_{c\bar{s}0}^a(2900)$, where the channels are denoted as $|1\rangle\equiv|D_s\pi\rangle$ and $|2\rangle=|D^\ast K^\ast\rangle$. The potentials will be parameterized as follows:
\begin{eqnarray}\label{eq:potcc}
v&=&\left[\begin{array}{cc}
0 & v_{i}\\
v_{i} & v_{e}
\end{array}\right],
\end{eqnarray}
where $v_i$ and $v_e$ represent the inelastic and elastic couplings, i.e., $v_i\equiv\langle D_s\pi|\hat{V}_i|D^\ast K^\ast\rangle$ and $v_e\equiv \langle D^\ast K^\ast|\hat{V}_e|D^\ast K^\ast\rangle$. The inelastic transition $D^\ast K^\ast\to D_s\pi$ can happen via exchanging the $K$, $K^\ast$, and $K_1$ mesons. A contact term effectively approximates the contributions from $K^\ast$ and $K_1$ exchange, while the $K$ exchange presents a unique challenge: it introduces a three-body ($D^\ast K\pi$) cut in the inelastic transition potential since the exchanged $K$ meson can be on its mass shell. In this work, our primary focus is on capturing the gross features of the $D_s\pi$ invariant mass spectrum through permitting a virtual state pole, instead of achieving a meticulous fit to the experimental data. Therefore, for simplicity, we adopt a contact approximation. However, it is critical to emphasize that the three-body ($D^\ast K\pi$) cut emerges in both the $D^\ast K^\ast \to D_s\pi$ and $D^\ast K^\ast \to DK$ (via one-pion exchange) decays. Consequently, as the data in the $DK$ channel become available in experiments, a comprehensive investigation of the line-shapes of $T_{c\bar{s}0}^a(2900)$, incorporating the intricate three-body cut, will be imperative, mirroring similar efforts undertaken for the $Z_c(3900)$~\cite{Albaladejo:2015lob,Wang:2020dko,Chen:2023def}.

The coupling between $D_s$ and $\pi$ are set to be zero for simplicity considering their weak interactions at low energy\footnote{The $D_s\pi$ is analogous to the $\eta_c\pi$ or $J/\psi\pi$ channels of the charmoniumlike states. The low-energy coupling strength of $\eta_c(J/\psi)\pi$ is known to be tiny~\cite{Yokokawa:2006td}. Thus following the strategies as those in Refs.~\cite{Albaladejo:2015lob,Wang:2020dko,Meng:2020ihj}, the coupling for the $D_s\pi$ is set to be zero.}. The coupled-channel LSEs takes the similar form as that in Eq.~\eqref{eq:intLSE}, while the $v$, $t$ and $G$ will be the $2\times2$ matrices, such as $G=\text{diag}\{G_1,G_2\}$. The $G_1$ and $G_2$ represent the two-body propagators of the channel $|1\rangle$ and $|2\rangle$, respectively, which read
\begin{eqnarray}\label{eq:reGe}
G_i=\int_0^{\Lambda_i}\frac{q^{2}dq}{(2\pi)^{2}}\frac{\omega_{i1}+\omega_{i2}}{\omega_{i1}\omega_{i2}}\frac{1}{E^{2}-(\omega_{i1}+\omega_{i2})^{2}+i\epsilon},
\end{eqnarray}
where $\omega_{ik}=\sqrt{\bm{q}^2+m_{ik}^2}$ denotes the energy of the $k$th particle in the channel $|i\rangle$. We use the relativistic propagators for both the elastic and inelastic channels, since the $D_s\pi$ is far below the $D^\ast K^\ast$ threshold. This propagator can be reduced to the non-relativistic from~\eqref{eq:Ge} up to some constant factors. We still adopt the cutoff scheme to regularize the integral~\eqref{eq:reGe}. Because the momentum of $D_s\pi$ can approximatively reach up to $0.8$ GeV when the center of mass energy $E$ is at the $D^\ast K^\ast$ threshold, so we use a relatively hard cutoff for this channel, i.e., $\Lambda_1=1.0$ GeV. The soft cutoff $\Lambda_2=0.4$ GeV is used for keeping consistence with our calculations in Sec.~\ref{sec:nums}.

There are four Riemann sheets in the two-channel problem, and they will be denoted as $\{\zeta_1,\zeta_2\}$ with the following replacement for analytic continuation
\begin{eqnarray}\label{eq:anatic}
G_i(E+i\epsilon)\to G_i(E+i\epsilon)+\frac{i\zeta_i p_i}{4\pi E},
\end{eqnarray}
where
\begin{eqnarray}
p_i=\frac{\sqrt{[E^2-(m_{i1}+m_{i2})^2][E^2-(m_{i1}-m_{i2})^2]}}{2E}.
\end{eqnarray}
The $\zeta_i$ in Eq.~\eqref{eq:anatic} can be $0$ and $1$, so the four possible combinations correspond to four different Riemann sheets. The physical sheet is given as $\{0,0\}$ in our notation. According to the classification of poles in Ref.~\cite{Badalian:1981xj}, the virtual state pole in two-channel case resides in the $\{0,1\}$ sheet.

\begin{figure}[!ht]
\begin{centering}
    \scalebox{1.0}{\includegraphics[width=\linewidth]{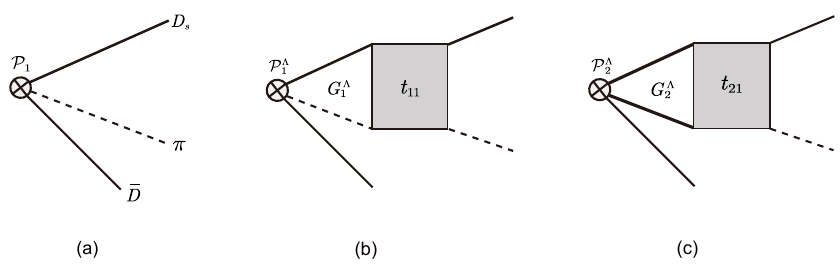}}
    \caption{Diagrams for the production process $B\to \bar{D}D_s\pi$, in which (a) denotes the direct production (no rescattering), while (b) and (c) represent the $D_s\pi$-driving ($\mathcal{P}_1^\Lambda$) and $D^\ast K^\ast$-driving ($\mathcal{P}_2^\Lambda$) productions accompanying the rescattering effect, respectively.\label{fig:production}}
\end{centering}
\end{figure}

We will simulate the production of $T_{c\bar{s}0}^a(2900)$ since it was observed in the process $B\to \bar{D}D_s\pi$~\cite{LHCb:2022sfr,LHCb:2022lzp}, and the corresponding Feynamn diagrams are shown in Fig.~\ref{fig:production}. From Fig.~\ref{fig:production}, the production amplitude can be written as~\cite{Dong:2020hxe,Wang:2020htx,Meng:2022wgl}
\begin{eqnarray}\label{eq:prodamp}
\mathcal{M}&=&\mathcal{P}_1+\mathcal{P}_1^\Lambda G_1^\Lambda t_{11}+\mathcal{P}_2^\Lambda G_2^\Lambda t_{21}\nonumber\\
&=&\mathcal{P}_1+(\mathcal{P}_1^\Lambda G_1^\Lambda v_{12}^\Lambda+\mathcal{P}_2^\Lambda)G_2^\Lambda t_{21}\nonumber\\
&=&\mathcal{P}_1+\mathcal{P}_2 t_{21},
\end{eqnarray}
in which we have used $t_{11}=v_{12}^\Lambda G_2^\Lambda t_{21}$ [note that the term $v_{11}^\Lambda(1+G_1^\Lambda t_{11})$ vanishes since $v_{11}^\Lambda=0$ in our case]. The $v_{ij}^\Lambda$ are the matrix elements of Eq.~\eqref{eq:potcc}, where the cutoff dependence is implied, while the $G_i^\Lambda$ correspond to the $G$ functions in Eq.~\eqref{eq:reGe}. One sees from the last line of Eq.~\eqref{eq:prodamp} that the first term only serves as a background, and the main contribution will come from the second term if there exists a near-threshold pole in the $t_{21}$. Thus, for the near-threshold invariant mass distribution, we can write it as
\begin{eqnarray}
\frac{d\Gamma}{dM_{D_s\pi}}\propto |t_{21}|^2 k_{s}k_{\pi}^\ast,
\end{eqnarray}
where $k_s$ is the momentum of the spectator $\bar{D}$ in the static frame of $B$ meson, and $k_{\pi}^\ast$ is the momentum of pion ($D_s$) in the center of mass system of $D_s\pi$.

We will produce a virtual state pole in the $\{0,1\}$ sheet with the potential in Eq.~\eqref{eq:potcc}, and the precondition $v_e\gg v_i$ is imposed. Because the virtual state is dominantly generated from the $D^\ast K^\ast$ interactions $v_e$, and the $D_s\pi$ should play the same role as the $J/\psi\pi$ for the $Z_c$ states (see the review~\cite{Meng:2022ozq}). The $v_i$ and $v_e$ are tuned to produce a virtual state pole that lies below the $D^\ast K^\ast$ threshold about $0.1$$-$$13$ MeV [corresponding to the upper and lower limits of $T_{c\bar{s}0}(2900)$ in Table~\ref{tab:DDbarandDKsys}] but with a small imaginary part\footnote{The opening of the $D_s\pi$ channel will give the pole an imaginary part. The decay width of $T_{c\bar{s}0}^a(2900)\to D_s\pi$ was calculated to be in the range $0.55$$-$$8.35$ MeV~\cite{Yue:2022mnf}, so we use a median $5$ MeV to tentatively study the line-shapes.}, i.e., $E_V=m_V+i\Gamma_V/2$, with $m_V\in[2887.6,2900.5]$ MeV and $\Gamma_V=5$ MeV.

We also investigate the effect of $K^\ast$ width considering $\Gamma_{K^\ast}\gg\Gamma_{D^\ast}$ (e.g., $\Gamma_{K^\ast}\approx50$ MeV, and $\Gamma_{D^\ast}\approx50$ keV~\cite{Workman:2022ynf}, so we will set $\Gamma_{D^\ast}=0$ in calculations), which is included in Eq.~\eqref{eq:reGe} via using a complex mass $m_{K^\ast}-i\Gamma_{K^\ast}/2$. In principle, the $\Gamma_{K^\ast}$ is a distribution with respect to its center of mass energy. Here, we treat it as a constant and consider its effective contributions. The effective width $\Gamma_{K^\ast}^{\text{eff}}$ is taken as three typical values, $0$, $25$, and $50$ MeV, respectively. 

\begin{figure*}[!ht]
\begin{centering}
    \scalebox{1.0}{\includegraphics[width=\linewidth]{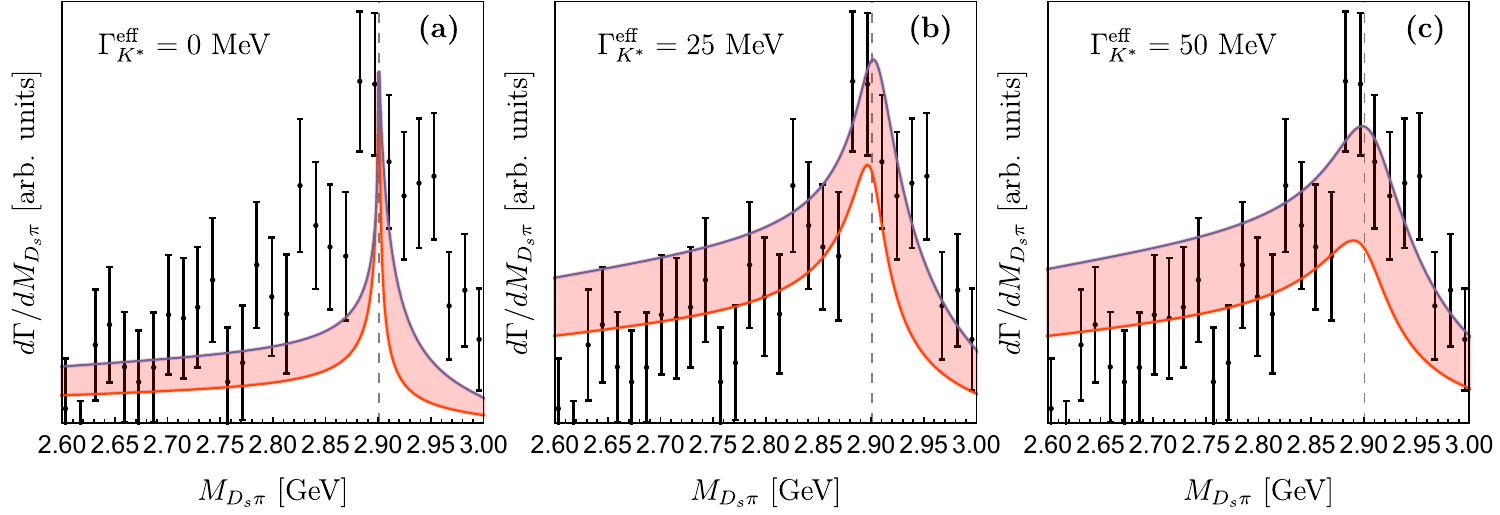}}
    \caption{The line-shapes of $d\Gamma/dM_{D_s\pi}$ in the virtual state cases and comparisons with the experimental events, in which the effective width of $K^\ast$ is taken as three typical values $0$, $25$ and $50$ MeV, respectively. The blue and red curves correspond to the poles that are below the $D^\ast K^\ast$ threshold about $0.1$ and $13$ MeV, respectively (the upper and lower limits in Table~\ref{tab:DDbarandDKsys}), and the results in this region are denoted as the red band. The vertical dashed line denotes the threshold of $D^\ast K^\ast$. The data are taken from Ref.~\cite{LHCb:2022sfr}, and the background contributions have been subtracted.\label{fig:lineshapes}}
\end{centering}
\end{figure*}

The invariant mass distributions of $D_s\pi$ and comparisons with the experimental events are shown in Fig.~\ref{fig:lineshapes}. In Fig.~\ref{fig:lineshapes} (a), one can see that it sharply peaks at the $D^\ast K^\ast$ threshold, and this is the generic feature of virtual states~\cite{Frazer:1964zz}. However, this sharp peak will be smoothed out when the finite width of $K^\ast$ is considered. The results in Figs.~\ref{fig:lineshapes} (c) and (d) both can qualitatively describe experimental data, which implies that the intrinsic width of $K^\ast$ plays significant role in shaping the $D_s\pi$ invariant mass spectrum (one can see Ref.~\cite{Frazer:1964zz} for more general discussions). The analyses of the line-shapes indicate that the molecular (virtual state) interpretation of $T_{c\bar{s}0}^a(2900)$ is consistent with the experimental data.

In Sec.~\ref{sec:spec}, we have predicted the partners of $T_{cs0}(2900)$ and $T_{c\bar{s}0}^a(2900)$, and some of them can be searched for in the $\bar{D}K$, $\bar{D}^\ast K$ and $DK$, $D^\ast K$, $D_s^\ast \pi$ channels, respectively. If these states indeed exist, one can expect the similar line-shapes near the $\bar{D}^{(\ast)} K^\ast$ and $D^{(\ast)}K^\ast$ thresholds in the corresponding inelastic channels.


\section{Summary and outlook}\label{sec:sum}

We have proposed an effective potential model to describe the low-energy residual strong interactions between heavy hadrons. We notice that the typical strength of the near-threshold interactions is too weak to excite the strange quarks inside the hadrons, which implies that the strange quark will behave as a {\it heavy} quark (inactive source) for the near-threshold interactions.
The interactions are primarily governed by the interplay between the light $u$ and $d$ quarks in two separate {\it heavy} hadrons. This forms the basis for linking the $\bar{D}^{(\ast)} K^\ast$ and $D^{(\ast)} K^\ast$ systems to the $D^{(\ast)}D^{(\ast)}$ and $D^{(\ast)}\bar{D}^{(\ast)}$ systems, respectively, based on their interaction types.

In the molecular picture, we use the three $P_c$ states, $T_{cc}(3875)$, and $X(3872)$, $Z_c(3900)$ to determine the LECs of the $qq$ and $\bar{q}q$ interactions, respectively. The existence of bound/virtual states in the heavy-(anti)heavy sectors will naturally lead to the emergence of virtual states in the $\bar{D}^{(\ast)} K^\ast$ and $D^{(\ast)} K^\ast$ systems. The recently observed $T_{cs0}(2900)$ and $T_{c\bar{s}0}^a(2900)$ can be well identified as the charmed strange partners of the $T_{cc}(3875)$ and $Z_c(3900)$, respectively. Besides, we also predict many members as the good molecule candidates in our systematic studies, such as the $0(1^+)$ $D^\ast D^\ast$, $0(1^+)$ $\bar{D}K^\ast$, $0(1^+)$ $\bar{D}^\ast K^\ast$, $0(2^+)$ $\bar{D}^\ast K^\ast$, and $0^+(2^{++})$ $D^\ast\bar{D}^\ast$, $1^-(0^{++})$ $D^\ast\bar{D}^\ast$, $0(2^+)$ $D^\ast K^\ast$, $1(1^+)$ $D^\ast K^\ast$, while the $0^+(0^{++})$ $D\bar{D}$ and $0(1^+)$ $DK^\ast$ might also exist as the virtual states. We suggest to use the $0(1^+)$ $DK^\ast$ as a key to infer the existence of $0^+(0^{++})$ $D\bar{D}$ since they share the same effective potential.

We have also analyzed the experimental line-shapes of $T_{c\bar{s}0}^a(2900)$ in a coupled-channel framework, in which we consider two channels, $D_s\pi$ and $D^\ast K^\ast$. We can qualitatively describe the $D_s\pi$ invariant mass spectrum in the molecular setup for $T_{c\bar{s}0}^a(2900)$ after considering the finite width of $K^\ast$. Thus, the molecular (virtual state) interpretation of $T_{c\bar{s}0}^a(2900)$ is consistent with the experimental data. Similar line-shapes in the elastic channels are expected for the other members in the $\bar{D}^{(\ast)} K^\ast$ and $D^{(\ast)} K^\ast$ systems.

Establishing the spectrum of hadronic molecules holds equal significance to constructing the spectrum of individual hadrons, as both are outcomes of strong interactions occurring at distinct energy scales.  There shall emerge a complete molecule spectrum if the mentioned states in this work are indeed the hadronic molecules. We suggest to search for the predicted states in experiments.

\section*{Acknowledgement}
This work is supported by the National Natural Science Foundation of China under Grants Nos. 12105072, 12305090, 11975033 and 12070131001. B. Wang was also supported by the Youth Funds of Hebei Province (No. A2021201027) and the Start-up Funds for Young Talents of Hebei University (No. 521100221021). This project is also funded by the Deutsche Forschungsgemeinschaft (DFG, German Research Foundation, Project ID 196253076-TRR 110).

\bibliography{refs}
\end{document}